\documentclass[conference]{IEEEtran}

\usepackage[nolist]{acronym}
\usepackage{amsmath}
\usepackage[table,xcdraw]{xcolor}
\usepackage{balance}
\usepackage{caption}
\usepackage[letterpaper, left=1in, right=1in, bottom=1in, top=0.75in]{geometry} 
\ifCLASSINFOpdf
\usepackage[pdftex]{graphicx}
\else
\fi
\begin{document}
\bstctlcite{IEEEexample:BSTcontrol}
\title{Resilience of airborne networks}

\author{\IEEEauthorblockN{Hamed Ahmadi\IEEEauthorrefmark{1}\IEEEauthorrefmark{2}, Gianluca Fontanesi\IEEEauthorrefmark{1}, Konstantinos Katzis\IEEEauthorrefmark{3}, Muhammad Zeeshan Shakir\IEEEauthorrefmark{4},\\ and Anding Zhu\IEEEauthorrefmark{1}}
\\
\IEEEauthorrefmark{1} School of Electrical and Electronic Engineering, University College Dublin, Ireland.\\
\IEEEauthorrefmark{2} School of Computer Science and Electronic Engineering, University of Essex, UK\\
\IEEEauthorrefmark{3} Department of Computer Science and Engineering, European University Cyprus, Cyprus\\
\IEEEauthorrefmark{4} School of Engineering and Computing, University of the West of Scotland, Paisley, Scotland, UK\\
%\thanks{This publication has emanated from research conducted with the financial support of Irish Research Council under Grant GOIPG/2017/1741.}

}

\maketitle
\vspace*{-12mm}
\begin{abstract}
%5G networks are expected to be always available and highly reliable. 
%With recent advances in development of airships, drones and other flying platforms pro 
Networked flying platforms can be used to provide cellular coverage and capacity. Given that 5G and beyond networks are expected to be always available and highly reliable, resilience and reliability of these networks must be investigated. This paper introduces the specific features of airborne networks that influence their resilience. We then discuss how machine learning and blockchain technologies can enhance the resilience of networked flying platforms. 
\end{abstract}

\begin{IEEEkeywords}
Networked flying platforms, resilience, self-organizing networks, machine learning, blockchain
\end{IEEEkeywords}

\vspace*{-2mm}
\section{INTRODUCTION}
%\vspace*{-1mm}
\acp{UAV} introduced a new challenge to cellular networks by acting as flying \acp{UE} that have much higher elevation than ground users. In \ac{5G} and beyond networks \acp{UAV} will be used for providing network services to ground and flying \acp{UE}. In \cite{Ahmadi_NFP17}, we introduced an architecture for \ac{NFP} in \ac{5G} and beyond networks. In this work, we aim to introduce and investigate the issues related to the resilience of airborne networks and in particular the introduced \acp{NFP}. 
\vspace*{-4mm}
\section{Network Resilience}
%\vspace*{-2mm}
In the literature there are several definitions for resilience of networks. In \cite{alliance20155g} resilience is defined as the capability if the network to recover from the failures. Sterbenz et al in \cite{sterbenz2010resilience} define resilience as the ability of the network to provide and maintain an acceptable level of service in the face of various faults and challenges to normal operation. According to \cite{sterbenz2010resilience} resilience disciplines are classified into two classes of challenge tolerance and trustworthiness related disciplines. The first class relate to the design of the system and include survivability, disruption tolerance and traffic tolerance. The class of trustworthiness disciplines relate to system performance and include dependability, security and performability.
%
%With respect to wireless networks \cite{gomez2014enabling} introduces three factors that limit cellular networks’ resilience in disastrous areas. These factors are physical destruction of network components, strong dependence between user equipment (UE) and access network, and evolved packet core (EPC) breakdown. 
%
In studying the resilience of airborne networks, we follow the definition provided by \cite{sterbenz2010resilience} and consider all the mentioned disciplines. However, due to the special case of airborne networks we need to emphasize on some of the disciplines and add new ones. This is mainly because \cite{sterbenz2010resilience} focuses on fixed networks and misses the features of wireless like multi-operator environments and spectrum/infrastructure sharing.
\vspace*{-2mm}
\section{\ac{NFP} features affecting the resilience}
\vspace*{-1mm}
Talking about resilience of each type of networks, we have to consider its and its components specifications and limitations. \acp{NFP} have unique features that affect their resilience. 
\vspace*{-4mm}
\subsection{Mobility} 
\vspace*{-1mm}
In an \ac{NFP}, e.g. 3-layer architecture in \cite{Ahmadi_NFP17}, all the \acp{HAP}, \acp{MAP} and \acp{LAP} are not fixed and have the ability to change their position and possibly their altitude. In one hand, mobility introduces challenges like possible collisions among the flying platforms, backhaul challenges, and connection loss. On the other hand, mobility enables the network to proactively respond to unpredicted events like \ac{UAV} failures or a sudden appearance of a demand hotspot. In the first scenario, the platforms especially \acp{LAP} can re-organise to preform self-healing, while in the second scenario an \ac{LAP} can move closer to the demand hotspot reducing the distance between the access point and \acp{UE} and the other \acp{LAP} reshape to cover the rest of the area. Although mobility is not considered in \cite{sterbenz2010resilience} classification, it will have a significant influence on challenge tolerance related disciplines like distribution tolerance and traffic tolerance as can be seen in the examples above.
\vspace*{-2mm}
\subsection{Energy limitations}
\vspace*{-1mm}
Most of the existing work \cite{Ahmadi_NFP17,naqvi2018drone} consider battery powered \acp{UAV} as the \acp{LAP}. This means that the \acp{UAV} have a limited operation time and need to fly to charging stations imposing a (predictable) disruption to the network. This feature clearly illustrates the importance of reliable self-organising mechanisms in the network. In these scenarios the self-organising system can either seamlessly replace the leaving \ac{UAV} with another \ac{UAV} (redundant), or change the network parameters, including the position of \acp{LAP}, to deal with this disruption.
\vspace*{-1mm}
\subsection{Physical vulnerabilities} 
Flying platforms are physically more vulnerable to accidental and intentional disruptions than fixed networks components. Accidents that can take \acp{LAP} down include lightning, strong wind, and clashing with birds. Flying platforms can also be targets of intentional disruptions like shooting or spoofing. Moreover, intruding drones pretending to be members of \ac{NFP} can disrupt the network functionality without causing problem to a single platform. 
\vspace*{-1mm}
\subsection{Multi-operator environment}
Open air is not a restricted area, except restricted zones defined by authorities, and several \ac{NFP} operators and other professional/amateur drone operators can coexist. This dynamic environment will increase the chance of collisions, turbulences, interference, and line-of-sight blockage which 
affects both challenge tolerance and trustworthiness related features of the network.  
\subsection{secondary duties}
\begin{figure}
    \centering
    \includegraphics[width=0.8\columnwidth]{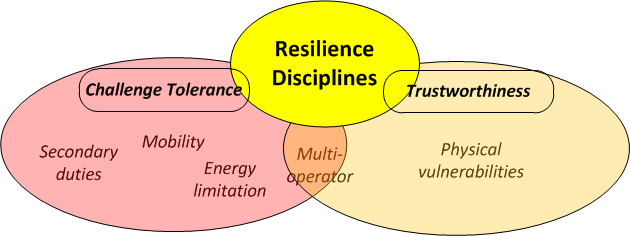}
    \setlength{\belowcaptionskip}{-6mm} 
    \caption{Classification of NFP features}
    \label{fig:SystemMOdel}
\end{figure}

According to design and need of the system flying platforms specially \acp{LAP} and \acp{MAP} can have secondary duties like surveillance or protecting the network by spoofing intruding \acp{UAV} \cite{naqvi2018drone}. Similar to energy limitation case, this may cause the \acp{UAV} to leave their network duties. Unlike the battery recharging case, the secondary duties are not always predictable, especially in the case of intrusion protection, which makes providing a redundant \ac{UAV} to seamlessly take on network duties of the leaving \ac{UAV} more challenging. In these scenarios self-healing mechanisms help the network to maintain its \ac{QoS}. Figure~\ref{fig:SystemMOdel} shows the classification of aforementioned features.  

\section{Strategies to enhance resilience of \acp{NFP}}

Most of the aforementioned features of \acp{NFP} that influence their resilience are classified as challenge tolerance-related which are not measurable. The challenges related to these features should be addressed in the design and engineering of the network. Although the resilience of a network cannot be measured based on its design and engineering, they effect dependability, security and performability of networks which are measurable.

The dynamic environment and the duties of \acp{NFP} require an architecture that enables autonomous reactions to different disruptions. Therefore, \ac{NFP} can benefit from \ac{SON} technologies. As defined in  \cite{valente2017SON_survey}, \acp{SON} are adaptive, autonomous, and they are able to independently decide when or how to trigger certain actions based on interaction with the environment. In \cite{Ahmadi_NFP17} we proposed a multi-layer architecture for \acp{NFP} and studied \ac{NFP} specific \ac{SON} features. To achieve better resilience for \acp{NFP} we can use machine learning and blockchain technologies.
\vspace*{-2mm}
\subsection{Machine learning}
\vspace*{-1mm}
A survey of machine learning techniques used in \ac{SON} for wireless networks and their applications is provided in \cite{valente2017SON_survey}. To the best of our knowledge there is no existing work that studies the application of \textit{machine learning} in self-organizing airborne networks. 

%A conventional static design will not work and provide resilience to \acp{NFP}.    \cite{valente2017SON_survey}

Resilinet project \cite{sterbenz2010resilience} proposes a two-phase resilience strategy where the first phase is responsible for dealing with the disruption and maintaining an acceptable level of service while the second phase aims to help the system to evolve and prevent and/or prepare for similar future disruptions. Phase one consists of detect, defend, remediate and recover, and phase two has two activities of diagnose and refine. A resilient airborne network can quickly detect disruptions and remediate. However, an \ac{NFP} can be more resilient using carefully trained learning algorithms that can predict disruptions like battery limitation or even possible intrusions.     

Optimization and game theoretic modeling are the most common methods in the existing works \cite{Ahmadi_NFP17} for planing the movement and position of flying platforms to maximize the coverage area or to maximize the delivered data rate to \acp{UE}. Several parameters effecting the decision of these algorithms which are traditionally set to an empirical mean value or inaccurately chosen can be learnt by machine learning algorithms based on the previous experiences \cite{ML5G}. This leads to faster and more accurate reaction to a disruption.
%Supervised learning can 
\vspace*{-2mm}
\subsection{Blockchain and smart contracts}
Blockchain can be defined as a resilient, reliable, transparent and decentralized way of storing and distributing a database across all nodes of a network \cite{malki2016automating}. Blockchain can assist with the security of \acp{NFP} against intruders pretending to be members of the network and/or spoofing attempts. 

In a multi-operator environment smart contracts can significantly help to manage space and spectrum sharing. A smart contract is basically a contract that its terms are enforced and executed automatically as computer codes among the participating entities without the need of an en-forcer or a third party. Smart contracts can facilitate deployment of automated charging stations at the roof of the building reducing the flight distance and time of \acp{UAV} to recharge their batteries. 
\vspace*{-2mm}
\section{Conclusions}
In this paper we introduced specific features of \acp{NFP} that affect their resilience. Most of these features are related to the design and engineering of the network, and are not easily measurable. We also named machine learning and blockchain as two promising technologies that can improve resilience of airborne networks. 
\vspace*{-2mm}
%\section*{Acknowledgement}
%\footnotesize This publication has emanated from research conducted with the financial support of Irish Research Council under Grant GOIPG/2017/1741.

% \ifCLASSOPTIONcaptionsoff
%   \newpage
% \fi

%\balance
\begin{acronym} 
\acro{5G}{Fifth Generation}
\acro{ACO}{Ant Colony Optimization}
\acro{BB}{Base Band}
\acro{BBU}{Base Band Unit}
\acro{BER}{Bit Error Rate}
\acro{BS}{Base Station}
\acro{BW}{bandwidth}
\acro{C-RAN}{Cloud Radio Access Networks}
\acro{CAPEX}{Capital Expenditure}
\acro{CoMP}{Coordinated Multipoint}
\acro{DAC}{Digital-to-Analog Converter}
\acro{DAS}{Distributed Antenna Systems}
\acro{DBA}{Dynamic Bandwidth Allocation}
\acro{DL}{Downlink}
\acro{FBMC}{Filterbank Multicarrier}
\acro{FEC}{Forward Error Correction}
\acro{FFR}{Fractional Frequency Reuse}
\acro{FSO}{Free Space Optics}
\acro{GA}{Genetic Algorithms}
\acro{HAP}{High Altitude Platform}
\acro{HL}{Higher Layer}
\acro{HARQ}{Hybrid-Automatic Repeat Request}
\acro{IoT}{Internet of Things}
\acro{LAN}{Local Area Network}
\acro{LAP}{Low Altitude Platform}
\acro{LL}{Lower Layer}
\acro{LOS}{Line of Sight}
\acro{LTE}{Long Term Evolution}
\acro{LTE-A}{Long Term Evolution Advanced}
\acro{MAC}{Medium Access Control}
\acro{MAP}{Medium Altitude Platform}
\acro{ML}{Medium Layer}
\acro{MME}{Mobility Management Entity}
\acro{mmWave}{millimeter Wave}
\acro{MIMO}{Multiple Input Multiple Output}
\acro{NFP}{Network Flying Platform}
\acro{NFPs}{Network Flying Platforms}
\acro{OFDM}{Orthogonal Frequency Division Multiplexing}
\acro{PAM}{Pulse Amplitude Modulation}
\acro{PAPR}{Peak-to-Average Power Ratio}
\acro{PGW}{Packet Gateway}
\acro{PHY}{physical layer}
\acro{PSO}{Particle Swarm Optimization}
\acro{QAM}{Quadrature Amplitude Modulation}
\acro{QoE}{Quality of Experience}
\acro{QoS}{Quality of Service}
\acro{QPSK}{Quadrature Phase Shift Keying}
\acro{RF}{Radio Frequency}
\acro{RN}{Remote Node}
\acro{RRH}{Remote Radio Head}
\acro{RRC}{Radio Resource Control}
\acro{RRU}{Remote Radio Unit}
\acro{SCBS}{Small Cell Base Station}
\acro{SDN}{Software Defined Network}
\acro{SNR}{Signal-to-Noise Ratio}
\acro{SON}{Self-organising Network}
\acro{TDD}{Time Division Duplex}
\acro{TD-LTE}{Time Division LTE}
\acro{TDM}{Time Division Multiplexing}
\acro{TDMA}{Time Division Multiple Access}
\acro{UE}{User Equipment}
\acro{UAV}{Unmanned Aerial Vehicle}

\end{acronym}

% Usage in text:
% \ac{BBU} for singular
% \acp{BBU} for plural

%\begin{thebibliography}{99}
\bibliographystyle{IEEEtran}
\bibliography{UAV_ref.bib}

%\end{thebibliography}

\end{document}